\documentclass[aps,prl,reprint,superscriptaddress]{revtex4-2}

\usepackage{graphicx}
\usepackage{upgreek}
\usepackage{amsmath,amsbsy}
\usepackage{comment} 
\usepackage{siunitx}
\usepackage[toc,page]{appendix}
\usepackage{amsmath,amssymb,amsfonts}
\usepackage{bbold}
\usepackage{xcolor}

\usepackage[colorlinks=true,linkcolor=black,citecolor=blue,urlcolor=blue]{hyperref}

\DeclareMathOperator{\Tr}{Tr}

\newcommand{\matrixel}[3]{\ensuremath{\langle #1 | #2 | #3 \rangle}}
\newcommand{\mean}[1]{\ensuremath{\langle #1 \rangle}}
\newcommand{\ket}[1]{\ensuremath{\left| #1 \right\rangle}}

\begin{document}

\author{Piero Naldesi}
\affiliation{Institute for Theoretical Physics, University of Innsbruck, Innsbruck A-6020, Austria}
\affiliation{Institute for Quantum Optics and Quantum Information of the Austrian Academy of Sciences, Innsbruck A-6020, Austria}

\author{Andreas Elben}
\affiliation{Institute for Quantum Information and Matter, Caltech, Pasadena, CA, USA}
\affiliation{Walter Burke Institute for Theoretical Physics, Caltech, Pasadena, CA, USA}
\affiliation{Institute for Theoretical Physics, University of Innsbruck, Innsbruck A-6020, Austria}
\affiliation{Institute for Quantum Optics and Quantum Information of the Austrian Academy of Sciences, Innsbruck A-6020, Austria}

\author{Anna Minguzzi}
\affiliation{Univ. Grenoble Alpes, CNRS, LPMMC, 38000 Grenoble, France}

\author{David Cl\'ement}
\affiliation{Université Paris-Saclay, Institut d’Optique Graduate School, CNRS, Laboratoire Charles Fabry, 91127, Palaiseau, France}

\author{Peter Zoller}
\affiliation{Institute for Theoretical Physics, University of Innsbruck, Innsbruck A-6020, Austria}
\affiliation{Institute for Quantum Optics and Quantum Information of the Austrian Academy of Sciences, Innsbruck A-6020, Austria}

\author{Beno\^it Vermersch}
\affiliation{Institute for Theoretical Physics, University of Innsbruck, Innsbruck A-6020, Austria}
\affiliation{Institute for Quantum Optics and Quantum Information of the Austrian Academy of Sciences, Innsbruck A-6020, Austria}
\affiliation{Univ. Grenoble Alpes, CNRS, LPMMC, 38000 Grenoble, France}

 \begin{abstract}
We provide a measurement  protocol to estimate 2- and 4-point fermionic correlations in ultra-cold atom experiments.
Our approach is based on combining random atomic beam splitter operations, which can be realized with programmable optical landscapes, with high-resolution imaging systems such as quantum gas microscopes.
We illustrate our results in the context of the variational quantum eigensolver algorithm for solving quantum chemistry problems. 
\end{abstract}

\title{Fermionic correlation functions from randomized measurements in programmable atomic quantum devices}


\maketitle

Traditionally, quantum algorithms are run with quantum computers made of qubits.
Another interesting possibility consists in using \emph{fermionic} quantum computers with fermions as elementary constituents~\cite{bravyi2002}.
These devices are in particular relevant for running fermionic quantum algorithms, without the technical overhead of representing fermions with qubits e.g., via a Jordan-Wigner transformation~\cite{Jordan1928}. 
Fermionic quantum algorithms can be used to solve numerous quantum problems.
This includes quantum chemistry ~\cite{Cao2019,Bauer2020,McArdle2020,ArguelloLuengo2019}, the quantum simulation of fermionic quantum states relevant to high energy physics~\cite{Banuls2020} and condensed matter~\cite{auerbach2012}. 
These applications stimulate efforts to engineer  fermionic quantum systems, in particular with ultracold atoms.
Using programmable optical lattices or tweezer arrays~\cite{Murmann2015, Kaufman2015,Bluvstein2022, Young2022, yan2022}, quantum gas microscopes~\cite{Bakr2009,Sherson2010,Weitenberg2011,Gross2021}, time-of-flight imaging systems~\cite{PhysRevLett.101.155303, Cayla2018}, one can indeed create, manipulate and measure fermionic quantum states at high fidelity, with single-site control.
However, in order to employ setups with ultracold atoms as a \emph{fermionic quantum processor} for running fermionic quantum algorithms, there is a significant challenge to tackle: the measurement of multipoint correlations that represent the result of the computation.
Here, we provide a measurement protocol based on randomized measurements~\cite{Elben2022} to
access such correlations, which can be implemented in
existing experimental setups.
For concreteness,
we will illustrate our measurement  protocol in the context of the variational quantum eigensolver (VQE) algorithm
~\cite{Peruzzo2014, Kandala2017, Hempel2018, Arute2020}.
The protocol is however general and should also find applications in the context of the quantum simulation of Hubbard models with ultracold atoms~\cite{Gross2017}.

\begin{figure}[h!!!!t!]
 \centering
 \includegraphics[width=0.9 \columnwidth]{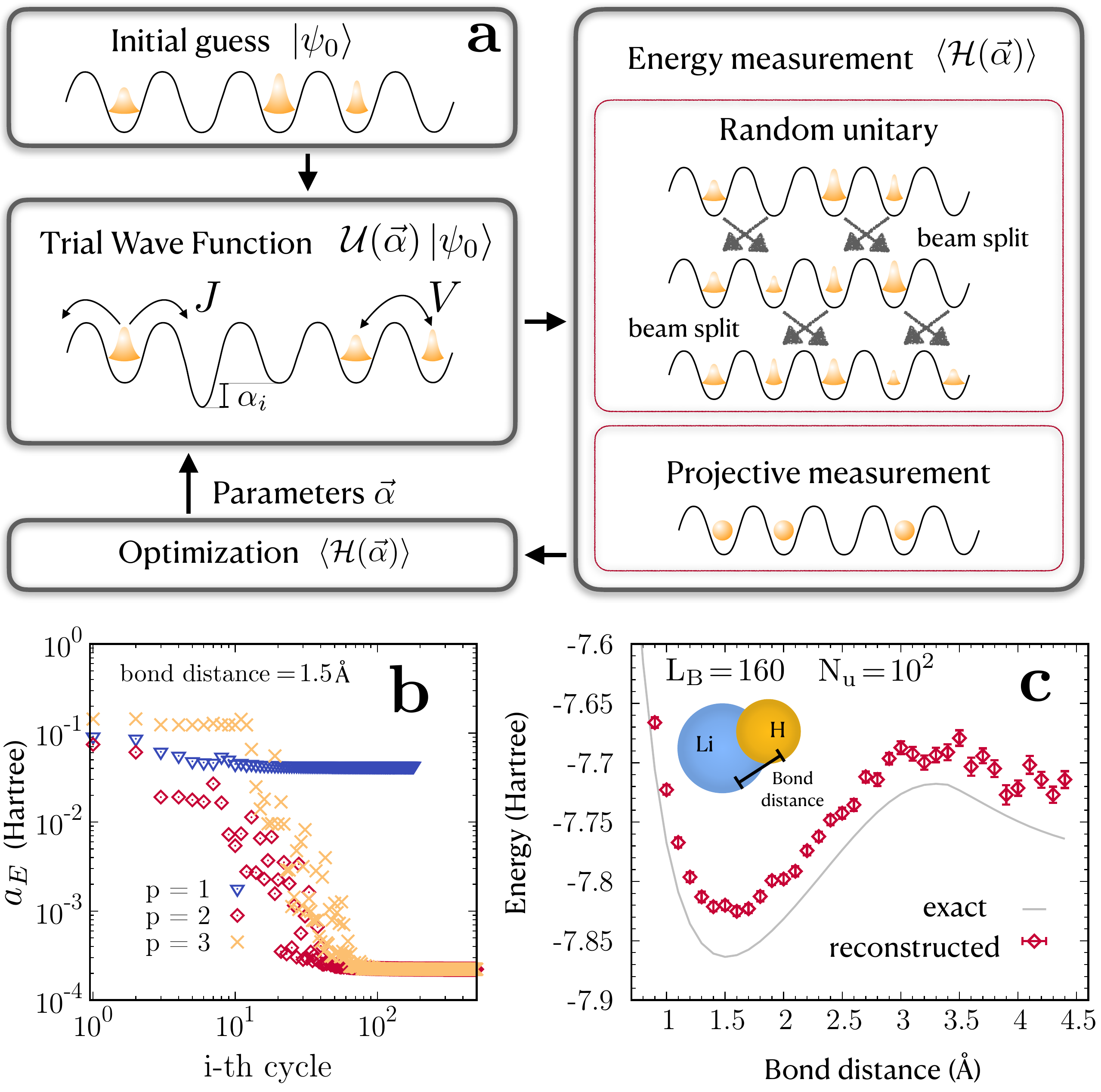}
\caption{ {\it Randomized measurement protocol for a fermionic variational quantum eigensolver (VQE).}
 \textbf{a}) VQE protocol with ultracold fermions in an optical lattice. The energy measurement is realized by random unitary transformations followed by projective measurements.
 \textbf{b}) Numerical study of VQE protocol for the Li-H molecule using the Hamiltonian Eq.~\eqref{eq:resource} ($V\!=\!J\!=\!1/T$). We plot the accuracy $a_E\!=\!|E_{VQE}\!-\!E|$, being $E$ the exact ground state energy and $E_{VQE}$ the one obtained from the VQE algorithm. Different depths $p$ have been considered.
 \textbf{c}) Reconstruction of the ground state energy as a function of the bond distance for the Li-H molecule. $N_u=10^2$ random unitary transformations have been considered (with an infinite number of projective measurements per unitary $N_m=\infty$).}
 \label{fig:vqe}
\end{figure}

VQE is a hybrid classical-quantum algorithm whose aim is to access the ground state of a quantum chemistry Hamiltonian.
As we explain below, a crucial step in this algorithm is the measurement of the expectation value of the molecular Hamiltonian, which can be expressed as a linear combination of the 2- and 4-point fermionic correlation functions
\begin{align}
C^{(1)}_{ij} = \mean{c^{\dagger}_i c_j}
\ \ \text{and} \ \ 
C^{(2)}_{ijkl} = \mean{c^{\dagger}_i c_j c^{\dagger}_k c_l}.
\label{eq:corr}
\end{align}
where the $c^\dagger_i$ ($c_i$) are fermionic creation (annihilation respectively) operators.
While several approaches exist in the literature for the measurement of the $C^{(1)}_{ij}$ matrix \cite{Knap2013, Brown2020, Gluza2021}, the estimation of the four point correlator $C^{(2)}_{ijkl}$ is a key point addressed by our work.

Our measurement protocol takes advantage of existing high-resolution imaging systems, such as  quantum gas microscopes, or single-atom-resolved detection methods after a time-of-flight. These methods provide snapshots of the fermionic populations in a given basis (position space or momentum space), giving access to `diagonal' correlations of the type $N_i=\mean{c^\dag_ic_i}$ and $N_{ij}=\mean{c^\dag_i c_i c^\dag_j c_j}$. 
In order to access off-diagonal elements in Eq.~\eqref{eq:corr}, we propose to use atomic beam splitters \cite{Murmann2015}. This well-established technique effectively realizes a linear transformation of the $L$ modes of the type $ \mathbf{c}^{(U)} \equiv U.\mathbf{c} \!=\! (\sum_{j=1}^L U_{ij}c_j)_{i=1, \dots, L}$, where in our case $U$ is chosen as a random unitary matrix.
In the following we show that measurements performed in a fixed basis, \emph{after} application of $U$, can be mapped to the desired correlations Eq.~\eqref{eq:corr}. This approach is adapted from the concept of randomized measurement protocols, now routinely used in qubit-type experiments \cite{Elben2022}
(see in particular  Ref.~\cite{Zhao2021} for accessing fermionic Hamiltonians in qubit-based quantum computers).
In contrast to previous work for fermionic systems based on estimating the density matrix~\cite{Ohliger_2013} or entanglement entropies~\cite{Elben_hub, Vermersch2018}, our protocol accesses fermionic correlations with non-interacting unitary transformations,
and with a number of measurements that scales polynomially in system size, see below.
The use of randomized measurement allows us to derive analytically efficient estimators for the desired correlations and to study numerically the required number of measurements to reach a given statistical accuracy.

{\it The variational quantum eigensolver (VQE) with fermionic atoms---}
In order to introduce our measurement protocol, we find it instructive to 
recall the basic steps of VQE, and to illustrate the algorithm with ultracold atoms.   
The setup we have in mind is depicted in Fig.~\ref{fig:vqe}a).
The VQE algorithm has been introduced in the context of quantum chemistry to study the electronic ground state of a molecule~\cite{Peruzzo2014}. The molecular Hamiltonian is first cast in a second quantization form, see Supplementary Material (SM)~\cite{SM} 
\nocite{Cao2019,baier2016extended,chomaz2022,Islam2015}
%
\begin{align}
\mathcal{H}(\mathbf{R}) = 
\sum_{i, j=1}^{L} 
h^{(1)}_{ij} (\mathbf{R})
c^{\dagger}_i c_j + 
\sum_{i,j,k,l=1}^{L} 
h^{(2)}_{ijkl} (\mathbf{R})
c^{\dagger}_i c_j c_k^\dag c_l.
\label{genham}
\end{align}
The operators $c_j$ ($c_j^\dag$) are fermionic annihilation (creation) operators that describe a set of electronic orbitals. The coefficients $h^{(1)}_{ij} (\mathbf{R})$ and $h^{(2)}_{ijkl} (\mathbf{R})$, both explicitly depending of $\mathbf{R}$, encode the geometrical structure of the molecule, see SM~\cite{SM} for details.
Since $\mathcal{H}(\mathbf{R})$ is particle number conserving, we work in a fixed sector with $N$ spinless fermionic atoms. 
These $N$ particles, placed in a one-dimensional optical lattice made of  $L$ sites, correspond to the $N$ electrons in $L$ electronic orbitals of the original molecule.
Spinful fermions, or different geometries can be also used~\cite{Mandel2003, Mandel2003PRL, karski2009, Daley2008, Daley2011}.
The first term in $\mathcal{H}$ accounts for the single-electron problem of the molecule, i.e the kinetic energy of the electrons and the interactions with the nuclei. The second term represents the Coulomb interactions between electrons.
Due to this second term, finding the ground state of $\mathcal{H}$ 
is a challenging many-body problem in a Hilbert space growing exponentially with number of orbitals L.  

In the VQE, one first parametrizes a set of variational wave functions $\ket{\psi(\mathbf{\alpha})}$, where $\mathbf{\alpha}$ is a vector of adjustable parameters. Note that, in order to generate these wave functions, it is not required to physically implement the molecular Hamiltonian $\mathcal{H}$.
However, one should be able to measure the expectation value $\mean{\mathcal{H}(\mathbf{R})}_\alpha=\matrixel{\psi(\mathbf{\alpha})}{\mathcal{H}(\mathbf{R})}{\psi(\mathbf{\alpha})}$, which by linearity, and using fermionic anti-commutation relations, can be achieved by accessing all correlations $C^{(1)}_{ij}$ and $C^{(2)}_{ijkl}$.
The estimation $\mean{\mathcal{H}(\mathbf{R})}_\alpha$, obtained on the quantum system, is then used as input for a minimization routine executed on a classical computer.
This routine adjusts iteratively the parameters $\mathbf{\alpha}$ in order to minimize the cost function $\mean{\mathcal{H}(\mathbf{R})}_\alpha$.  

We show in Fig.~\ref{fig:vqe}b) a numerical illustration of the VQE optimization \emph{with ultracold atoms}. In our example of the Li-H molecule, we consider $N\!=\!2$, $L\!=\!4$. For a diatomic molecule, $\mathbf{R}$ reduces to the bond-distance $R$, i.e. the distance between the two atoms. We use standard numerical routines to obtain the parameters $h^{(1)}_{ij}(R)$ and $h^{(2)}_{ijkl}(R)$ that describe the molecular Hamiltonian for each value of $R$~\cite{McClean2020}.
The variational wave function $\ket{\psi(\mathbf{\alpha})}=e^{-iH_pT}\dots e^{-iH_1T}\ket{\psi_0}$ is then generated by applying a sequence of $s=1, \dots, p$ time evolutions  of duration $T$ ($\hbar=1$) to the initial state $\ket{\psi_0}$. Each of these quenches is driven by the extended Fermi-Hubbard model Hamiltonian 
\begin{align}
 H_s=\sum_{i=1}^L \left[-J(c^\dagger_{i+1}c_i+\text{h.c.})+\alpha_{i, s}n_i+Vn_i(n_{i+1})\right], \label{eq:resource}
\end{align} 
where the list $\mathbf{\alpha}=\alpha_{i, s}$ of $Lp$ spatial and time-dependent energy potentials represent the variational parameters to be optimized.
Here, $V$ represents interactions between neighboring sites that can be obtained for instance using dipolar interactions with magnetic atoms \cite{Ni2010, baier2016extended, chomaz2022}.
Fig.~\ref{fig:vqe}b) shows the convergence of VQE algorithm: At low depth $p\!=\!2$ and after a number of  optimization cycles $\sim10^2$, the algorithms reaches with high fidelity the ground state of the molecular Hamiltonian $\mathcal{H}$.

The success of the VQE optimizations relies on a precise estimate of $\mean{\mathcal{H}(\mathbf{R})}_\alpha$, and therefore of the full correlators $C^{(1)}_{ij}$ and $C^{(2)}_{ijkl}$.
We now present our measurements protocol giving access to such correlations.
As a first illustration, Fig~\ref{fig:vqe}c) shows a simulation of the measurement of the bond dissociation curve of Li-H, representing the molecular energy after the final iteration of the VQE optimization as a function of the bond distance $R$.

{\it Presenting our measurement protocol---}
We consider two options for implementing the measurement protocol. The first option is conceptually the simplest one but requires single-site addressing and imaging, e.g, using a quantum gas microscope.
The second option is tailored instead for `time-of-flight' experiments, and replaces the measurement of single site population $N_i$ and correlations $N_{i,j}$ by the populations $N_k$ and correlations $N_{k,k'}$ between different momentum components.
For the first option, our measurement protocol begins by making the system non-interacting, e.g, by changing the dipole moment of the atoms via a change of internal levels.
This step is not required for the second option, see details below. 
Our measurement protocol then consists in applying successively a sequence of two-site random beam splitter operations to create a global random transformation. Two-site beam splitter operations are engineered between two adjacent lattice sites $i,i+1$, e.g by ramping potential barriers \cite{Islam2015} and can be realized with high fidelity in present experimental setups \cite{Murmann2015,Islam2015}. For each two-site beam splitter operations, the system then evolves according to a two-site free-fermionic Hamiltonian with the lattice operators being transformed in the Heisenberg picture as
\begin{align}
\begin{pmatrix}
 c_i\\ c_{i+1}
 \end{pmatrix}\to 
 u_i(\alpha,\phi,\psi)
\begin{pmatrix}
c_i\\c_{i+1} 
\end{pmatrix}.
\label{eq:beamsplitter}
\end{align}
Here, $u_i(\alpha,\phi,\psi)$ is a $2\times 2$ unitary matrix which can be parametrized by three angles $\alpha,\psi \in [0,2\pi]$ and $\phi \in [-\pi/2,\pi/2]$ (see SM~\cite{SM} for more details).  A global unitary transformation from the circular unitary ensemble (CUE)~\cite{Mezzadri2007} is then generated by a sequence of $L(L-1)/2$ such beam splitters
\mbox{$U=\prod_{j=1}^{L-1}\prod_{i=1}^{j} \mathbb{1}_{i} \otimes u_{i}(\alpha_{j},\phi_{j,i},\psi_{j,i})\otimes \mathbb{1}_{L-i-1}$} where the angles $\alpha_{j},\phi_{j,i},\psi_{j,i}$  are sampled independently from particular distributions (see SM~\cite{SM} and the schematic Fig.~\ref{fig:vqe} for more details).
The same techniques can be extended to spin-dependent lattices \cite{jaksch1999entanglement, Mandel2003, Daley2008}, allowing for the implementation of beam splitters in spin-full fermion systems.

At the end of the sequence described above, the Heisenberg operators c are transformed to $ \mathbf{c}^{(U)} \!=\! U.\mathbf{c}$ , and a projective measurement is realized.
By measuring the occupation of the lattice sites, we obtain estimates of any expectation value of the form $N^{(U)}_i\!\!\!=\!\!\mean{ n^{(U)}_i}$ and $N^{(U)}_{ij}\!=\!\mean{ n^{(U)}_i n^{(U)}_j}$.
This procedure (sequence of random beam splitters and projective measurement) is repeated for a number $N_u$ of random transformations. 
As we show now, the statistics of these measurements yield the correlations in Eq.~\eqref{eq:corr}.

{\it Extracting correlations from the measured data --- }
In order to extract $C^{(1)}_{ij}$ from the  measured data $N_i^{(U)}$, the key idea in randomized measurements consists in using the fact that statistical correlations of random unitary transformations $U$ can be calculated using the theory of unitary $n$-designs \cite{Collins2003,Roberts2017}. As shown in the SM~\cite{SM}, we obtain
\begin{align}
C^{(1)}_{ij} &=
 L \sum_{s_1,s_2=1}^L
(-L)^{\delta_{s_1,s_2}-1} \: 
\overline{
N^{(U)}_{s_2}
U_{s_1,i} U^*_{s_1,j}},
\label{1body}
\end{align}
where the overline denotes the average over many different unitary transformations drawn randomly from the CUE. Here, we have used the statistical correlations between $4$ matrix elements of $U$ and used accordingly the $2$-design properties of the CUE. 

Similarly, the 4-point correlation tensor $C^{(2)}$ is obtained from higher-order statistical correlations. To provide an analytical estimator, we need to consider a slight modification of our protocol: The system of $L$ sites is embedded in a larger optical lattice of $L_B$ sites with $L_B>L$, these additional $L_B\!-\!L$ sites are not occupied at the beginning of the measurement sequence.
The reason for this embedding is that in the limit $L_B\gg 1$, the statistical correlations of the large $L_B\times L_B$ matrices $U$ simplify drastically: Restricting indices to a subset of values, the matrix elements $U_{ij}$ become effectively independent Gaussian random variables, see \cite{Enk2012} and SM~\cite{SM}.
This allows us to invert the relation between the measured data and the correlations, and to write
\begin{align}
C^{(2)}_{ijkl} \! &= \!
\sum_{\mathbf{s}=1}^{L_B} \!
o^{(2)}_{\mathbf{s}} \overline{
N^{(U)}_{s_3,s_4} \:
U_{s_1,i} U^*_{s_1,j} U_{s_2,k} U^*_{s_2,l}}
+ \mathcal{O} \left( \frac{L^2}{L_B} \right)
\label{2body}
\end{align}
with $\mathbf{s}=s_1,s_2,s_3,s_4$, and 
\begin{align}
o^{(2)}_{\mathbf{s}}\! &= 
\frac{1}{L_B (L_B\!+\!3)!}
\begin{cases} 
  (L_B\!-\!4)! \! & \!  s_1 \! \neq \! s_2 \! \neq \! s_3 \! \neq \! s_4 \\[2pt]
- (L_B\!-\!3)! \! & \! (s_1 \! = \! s_3) \neq \!  s_2 \! \neq \!  s_4 \\[1pt]
- (L_B\!-\!3)! \! & \! (s_2 \! = \! s_4) \neq \!  s_1 \! \neq \!  s_3 \\[2pt]
  (L_B\!-\!2)! \! & \! (s_1 \! = \! s_3) \neq (s_2 \! = \! s_4)\\[2pt]
0  \! & \!  \text{otherwise}.
\end{cases}
\label{coeff2body}
\end{align}
The second term in Eq.~\eqref{2body} represents the non-Gaussian effects of the CUE matrices, that vanishes in the limit $L_B\to\infty$. Since the embedding procedure does not affect the reconstruction of $C_{ij}^{(1)}$, both correlation matrices can be reconstructed using the same experimental data. 

{\it Protocol with a time-of-flight apparatus---}
We now discuss the second option associated with ‘time-of-flight’ experiments, where we propose to implement beam splitters between momentum states during a time-of-flight expansion. 
Here, we consider that the fermionic state is prepared using the same resources, i.e the Fermi-Hubbard model Eq.~\eqref{eq:resource}. The measurement sequence is instead changed.
In this second scenario, ultracold atoms are first released from the optical lattices by switching off abruptly the latter. In our regime of interest for which the lattice filling is of order unity (or less), the expansion of the gas is driven by the zero-point energy in a lattice site and it is ballistic to an excellent approximation \cite{Cayla2018, Tenart2020}. After a short duration (set by the inverse of the zero-point energy in a site), the gas is diluted and exhibits the quadratic dispersion $\epsilon(k)$ of non-interacting atoms. Two-photon (Bragg) transitions then allow one to realise a beam splitter operation analogous to Eq.~\eqref{eq:beamsplitter} between two momentum states \cite{Lopes2015},
\begin{align}
\begin{pmatrix}
 \tilde c_k\\ \tilde c_{k+1}
 \end{pmatrix}\to 
 u_k(\alpha,\phi,\psi)
\begin{pmatrix}
\tilde c_k\\\tilde c_{k+1}
\end{pmatrix}
\label{eq:beamsplitter-kspace}
\end{align}
where $\tilde c_{k}=\sum_{j} c_{j} e^{-i 2\pi k j /L_B}/\sqrt{L_B}$, $k=0,\dots,L-1$, are the momentum-space operators. The energy difference $\Delta \omega$ between the two Bragg beams is set to match the energy difference $\epsilon(k+1)-\epsilon(k)$ of the two coupled momentum states. Implementing many beam splitters at once is permitted by the non-linearity of $\epsilon(k)$ and realised with multiple Bragg beams of varying energy differences $\Delta \omega$ \cite{Gadway2015}. Similarly to the first scenario, measuring the occupation of the momentum lattice sites yields expectation values of the form $\tilde N_k = \mean{\tilde n^{(U)}_k}$, $\tilde N_{k,k'} = \mean{\tilde n^{(U)}_k\tilde n^{(U)}_{k'}}$, from which correlators of the momentum operators are estimated using Eq.~\eqref{1body}-\eqref{2body}. Single-atom-resolved detection in time-of-flight experiments \cite{Jeltes2007, Bergschneider2018} is perfectly suited to this aim. Finally, Fourier transforms relate the correlations Eq.~\eqref{eq:corr} in the position space to those measured after expansion, in the momentum space.

\begin{figure}[h!!!!!!]
 \centering
 \includegraphics[width=\columnwidth]{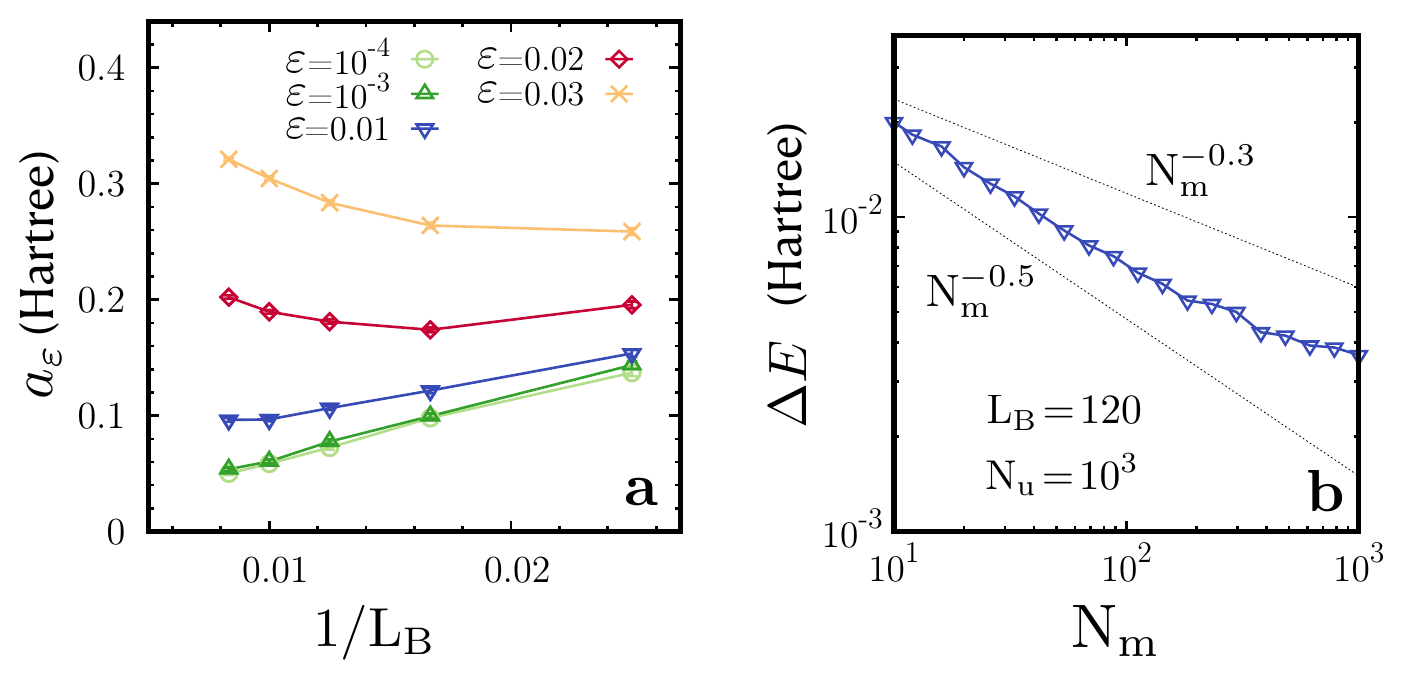}
 \caption{{\it Analysis of systematic and statistical errors.} \textbf{a}: Reconstruction error $a_\varepsilon$ due to unitary miscalibration as a function of $L_B^{-1}$  for different $\varepsilon$. Here, we consider $N_m =\infty$.
\textbf{b}: Scaling of statistical error $\Delta E$ as a function of number of measurements $N_m$. In both panels data refer to the Li-H molecule where the bond distance is fixed to 1.5 $\si{\angstrom}$ and $N_u=10^3$ unitary transformations have been considered.
 }
 \label{fig:misc}
\end{figure}

{\it Systematic and statistical errors---}
Finally, we address quantitatively the role of systematic and statistical errors in the measurement part of the protocol. Systematic errors can arise from the miscalibration of the random unitary transformations.
This effect, analyzed for instance for other randomized measurement protocols~\cite{Elben_cross}, is assessed here by adding random offsets to the three angles $\{ \alpha,\phi,\psi \}$ defining every beam splitter:
\begin{align}
\left\{ \alpha,\phi,\psi \right\} \rightarrow 
\left\{
 \alpha + \varepsilon \nu^\alpha, 
 \phi   + \varepsilon \nu^\phi, 
 \psi   + \varepsilon \nu^\psi
\right\},
\end{align}
where $\varepsilon$ tunes the miscalibration and the variables $\{ \nu^\alpha, \nu^\phi, \nu^\psi \}$ are picked uniformly from the interval $[-1,1]$.
A numerical study is shown in Fig.~\ref{fig:misc}a for various values of the miscalibration noise $\varepsilon$. The accuracy in the energy estimation is defined as $a_{\varepsilon} = |E_\varepsilon-E|$, with $E$ and $E_{\varepsilon}$ being respectively the exact energy and the one affected by miscalibration.
At large values of $\varepsilon$, the error is maximal at large $L_B$.
This is because the number of beam splitter operations increases with $L_B$, leading to the propagation of miscalibration errors.
Instead at small $\varepsilon$, we observe the influence of another source of systematic errors, which is the error in $O(1/L_B)$ appearing in the estimation of the 4-point correlations $C^{(2)}_{ijkl}$, c.f. Eq.~\eqref{2body}, see also additional numerical calculations in SM~\cite{SM}. This effect, responsible for the offset between reconstructed and exact result visible in Fig.~\ref{fig:vqe}c, can be reduced by enlarging $L_B$.
Finally, statistical errors arise from two contributions: (i) the finite number of random unitary transformations $N_u$ used to estimate the ensemble average in Eqs~\eqref{1body} and \eqref{2body}, (ii) the finite number of projective measurement $N_m$ used to obtain for a given unitary $U$, the expectation values $N^{(U)}_{s_1}$, and $N^{(U)}_{s_1,s_2}$.
In our protocol, the estimation of the ground state energy is obtained by averaging the results obtained from $N_u$ independently sampled unitary transformations. Thus, we quantify statistical errors in terms of the standard deviation on the mean $\Delta E=\delta E/\sqrt{N_u}$, with $\delta E$ is the standard deviation associated with estimations built from a single random unitary $N_u=1$.
Let us now address the role of the number of projective measurements $N_m$.
An important feature of the protocol is that the required value of $N_m$ to achieve a given accuracy in the measured correlations scales polynomially in system size $L_B$. This is because the quantities that need to be estimated via projective measurements in our protocol $N^{(U)}_{s_1}$, and $N^{(U)}_{s_1,s_2}$ are moments of random unitaries that scale as $1/L_B$ ($1/L_B^2$, respectively). 
As an illustration, we represent $\Delta E$ versus $N_m$ in Fig.~\ref{fig:misc}b for $L_B=120$, which also shows an approximate dependence in $\propto 1/\sqrt{N_m}$. In this case, we can for instance obtain a statistical error of order $\Delta E=0.1/\sqrt{N_u}$ for $N_m\approx 100$.
These results show that fermionic correlations can be estimated accurately via our protocol, and with a number of measurements that is compatible with the possibilities of existing experiments.

{\it Conclusion}---Our fermionic measurement protocol provides access to 2- and 4-point correlation functions in ultra-cold experiments. 
Beyond the practical use of this protocol for running fermionic quantum algorithms such as VQE in such quantum platforms, our work also points to several interesting research directions.
First, we can apply our ideas to higher order correlation functions, i.e beyond $C^{(2)}_{ijkl}$, in order for instance to measure Hamiltonian variances and check the convergence of the VQE algorithm.
Second, it will be also interesting to extend our protocol to continuous variable systems such as degenerate Fermi gases.

\begin{acknowledgments}
We thank J. Carrasco and C. Kokail for valuable discussions. Work in Innsbruck has been supported by the European Union's Horizon 2020 research and innovation programme under Grant Agreement No.\ 817482 (Pasquans), by the Simons Collaboration on Ultra-Quantum Matter, which is a grant from the Simons Foundation (651440, P.Z.), and by LASCEM via AFOSR No.\ 64896-PH-QC. A.E.\ acknowledges funding by the German National Academy of Sciences Leopoldina under the grant number LPDS 2021-02 and by the Walter Burke Institute for Theoretical Physics at Caltech. B.V. and A.M. acknowledges funding from the French National Research Agency (ANR-20-CE47-0005, JCJC project QRand). P.N. and B.V. acknowledge funding from the Austrian Science Foundation (FWF, P 32597 N).
D.C. acknowledges support from the Agence Nationale pour la Recherche (Grant Number ANR-17-CE30-0020-01) and the ``Fondation d'entreprise iXcore pour la recherche."
\end{acknowledgments}

\bibliography{main.bib}

\pagebreak
\clearpage

\section{Supplemental material}

In this supplemental material we provide further details and analytical proofs for the results we show in the main text. It is organized as follows: In Appendix A we show how to obtain the second quantization form of the molecular Hamiltonian \eqref{genham}. In Appendix B, we describe the Hurwitz decomposition for CUE matrices. In Appendix C we analytically derive Eqs.~\eqref{1body} and \eqref{2body} used to reconstruct 1- and 2-body correlation functions.  Last, in Appendix D, we describe the systematic error we introduce in the reconstruction of the 4-point correlations function by considering a finite $L_B$.

\section{Appendix A: Molecular Hamiltonian in second quantization}
\label{appendix:B}
In this section we show how to cast the molecular Hamiltonian in the form \eqref{genham}, from which the VQE algorithm can start.
We start from the first quantization formulation of the problem, in this frame the Hamiltonian of a molecule reads \cite{Tilly2021}:
\begin{align}
H_{1Q} &= - \sum_i \frac{\nabla_{\mathbf{R_i}}^2 }{2M_i} - \sum_i \frac{\nabla_{\mathbf{r_i}}^2}{2} - \sum_{i,j} \frac{Z_i}{|\mathbf{R_i} - \mathbf{r_j}|} \label{molham1q}
\\
&+ \sum_{i, j > i} \frac{Z_i Z_j}{|\mathbf{R_i} - \mathbf{R_j}|} + \sum_{i, j>i} \frac{1}{|\mathbf{r_i} - \mathbf{r_j}|}
\nonumber
\end{align}
Positions, masses, and charges of the nuclei are $\mathbf{R_i}, M_i, Z_i$. Positions of the electrons are $\mathbf{r_i}$. \\
Once an orbital basis for the electronic wave function is decided upon, under the Born-Oppenheimer approximation we can recast the Hamiltonian in the second quantized form:
\begin{align}
H_{2Q} = 
\sum^L_{ij=1} h^{(1)}_{ij} c^{\dagger}_i c_j + 
\sum^L_{ijkl=1} h^{(2)}_{ijkl} c^{\dagger}_i c^{\dagger}_j c_k c_l
\label{hamil}
\end{align}
where $\{ c^\dagger_i,c_j\}$ is a set of fermionic creation and annihilation operators. The number of sites, $L$, of this lattice Hamiltonian corresponds to twice the number of orbitals of the particular molecule.\\ 
The coefficients $h^{(1)}_{ij}$ and $h^{(2)}_{ijkl}$ can be obtained through:
\begin{align}
h^{(1)}_{ij} &= 
\int \ d\sigma \ \gamma_i^*(\sigma) \left(\frac{\nabla_\mathbf{r}^2}{2} - \sum_p \frac{Z_p}{|\mathbf{R_p} - \mathbf{r}|} \right)\gamma_j^*(\sigma) \label{chem_coef0} \\
h^{(2)}_{ijkl} &= 
\int \ d\sigma_1 \ d\sigma_2 \frac{\gamma_i^*(\sigma_1)\gamma_j^*(\sigma_2) \gamma_k(\sigma_1)\gamma_l(\sigma_2)}{2 |\mathbf{r_1} - \mathbf{r_2}|}. 
\label{chem_coef}
\end{align}
where the spatial and spin coordinates of the electrons are represented by the notation $\sigma_p=\{\mathbf{r_p},\sigma_p\}$ and $\gamma(\sigma)$ is a based form of spin dependent wave functions~\cite{Cao2019}. The coefficients in Eqs.~\eqref{chem_coef0} and ~\eqref{chem_coef} can be numerically evaluated for a broad range of molecules with, e.g., the help of the Python package OpenFermion \cite{openfermion}.

\section{Appendix B: Decomposition of $L\times L$ unitary transformations.}
\label{appendix:C}
In this appendix we explain how a $L\times L$ matrix of the circular unitary ensemble can be decomposed by a multiplication of $L(L\!-\!\ 1)/2$ SU(2) matrices, each of these matrices being parametrized by three Euler angles.
According to Refs.~\cite{Hurwitz1897, Diaconis2016}, any $L \times L$ unitary matrix can be written as a product of $L-1$ matrices $E_j$:
\begin{align}
U = e^{i \alpha_0} E_1 \dots E_{L-1},
\label{U}
\end{align}
%
where each of such matrices $E_j$ can in turn be expressed as: 
\begin{align}
E_j = 
u_{j,j} (\phi_{j,j}, \psi_{j,j}, & 0)
 u_{j,j-1} (\phi_{j,j-1},\psi_{j,j-1},0) \dots \nonumber \\
& u_{j,1} (\phi_{j,1},\psi_{j,1},\alpha_j) 
\label{Ej}
\end{align}
for a total of $L(L-1)/2$ multiplicative terms.
Each $u_{j,k}$  has a block diagonal structure and explicitly depends on three Euler angles $\psi_{j,k}$, $\alpha_{j}$ and $\phi_{j,k}$
\begin{align}
u_{j,k} \! &= \!
\begin{pmatrix}
\mathbb{1}_{k-1} \!\!\! & & &\\
 \!\!\! & e^{i \alpha_{j} } \cos (\phi_{j,k} ) & e^{i \psi_{j,k} } \sin (\phi_{j,k} ) & \\
 \!\!\! & -e^{-i\psi_{j,k} } \sin (\phi_{j,k} ) & e^{-i\alpha_{j} } \cos (\phi_{j,k} ) & \\
 \!\!\! & & & \!\!\! \mathbb{1}_{L-k-1}
\end{pmatrix}
\end{align}
with $\mathbb{1}_n$ denoting the identity matrix of size $n$.
Given the structure of the $E_j$ in equation \eqref{Ej}, and considering the random phase $\alpha_0$ in \eqref{U}, 
we end up with a total of $L^2$ independent parameters.
Any CUE matrix can be obtained by randomly selecting the three Euler angles of each $SU(2)$ in an appropriate way.
While the two angles $\psi_{j,k}$ and $\alpha_{j}$ must be uniformly chosen in the interval $(0,2\pi)$, $\phi_{j,k}$ is obtained from the $j$-dependent relation $\phi_{j,k} = \arcsin{\xi^{1/2j}_{j,k}}$, where the auxiliary variable $\xi_{j,k}$ is picked uniformly in the interval $(0,1)$ \cite{Diaconis2016}.

\section{Appendix C: Derivation of reconstruction formulas}

In this Appendix, we derive the reconstruction formulas Eqs.~\eqref{1body} and \eqref{2body} to estimate the full single- $C^{(1)}$ and two-body $C^{(2)}$ correlation matrices from randomized measurements of particle number density $N_i^{U}$ and particle number density-density correlators $N_{ij}^{U}$, respectively. Key ingredient is the Weingarten calculus of random unitary matrices \cite{Collins2003,Collins2010}. As explained in more details below, in the first case, estimation of $C^{(1)}$, we can make use of mathematical results obtained in Ref.~\cite{Elben2019}, essentially by replacing random unitary transformations acting on the full Hilbert space with transformations acting on position space. The second case, estimation of $C^{(2)}$, requires us to consider higher order Weingarten calculus. We show how how to reconstruct $C^{(2)}$ up to corrections that decrease with increasing system size. 

\subsection{Estimating the single-body correlation matrix}

Our aim is to prove Eq.~\eqref{1body} of the main text, relating the full single-body correlation matrix $C^{(1)}$ to correlations of randomized measurements of `diagonal' density correlators $N_i^{U}$. We first note that Eq.~\eqref{1body} is a relation of quantum expectation values, i.e.\ all quantum expectation values have been taken and all indices are spatial (lattice) indices: For a quantum state described by a density matrix $\rho$, the correlation matrix  $C^{(1)}\in \mathbb{C}^{L^2}$ with $L\times L$ matrix with complex entries
is defined as
\begin{align}
(C^{(1)})_{i,j=1,\dots, L} \equiv \Tr [ \rho \: c^\dagger_ic_j ].
\end{align}
In particular, $(N_i)_{i=1,\dots,L}=(\Tr [ \rho \: c^\dagger_ic_i ])_{i=1,\dots, L}$ is the vector consisting of its $L$ diagonal entries.

The space $\mathcal{M}_L \cong \mathbb{C}^{L^2}$ of complex $L\times L$ matrices forms a Hilbert space with inner product $(A,B)\equiv \Tr[A^\dagger B]$ for $A,B$. We denote the standard orthonormal basis $\{E_{ij}\}_{i,j=1,\dots L}$ where the matrices $E_{ij}$ have a single non-vanishing entry $(E_{ij})_{ij}=1$ and zeros otherwise. The operation $A \otimes B\in \mathcal{M}_L^{\otimes 2}$ denotes the standard Kronecker product of complex matrices.

To prove Eq.~\eqref{1body}, we start with the identity
\begin{align}
C^{(1)} &= \text{Tr}_2\left[ \mathbb{S} \: \mathbb{1}_L \otimes C^{(1)} \right]
\label{eq:swap_tom}
\end{align}
 where $\text{Tr}_2[A\otimes B] \equiv A \Tr[B]$ denotes the partial trace with respect to only the second `spatial copy' of the system. Further, we denote the identity matrix with $\mathbb{1}_{L}= \sum_{i=1}^L E_{ii}$ and define the `swap matrix'
\begin{align}
\mathbb{S} \equiv \sum_{i,j=1}^L E_{i,j} \otimes E_{j,i} \in \mathcal{M}_L^{\otimes 2}.
\end{align}
The identity \eqref{eq:swap_tom} can be readily proven by inserting the definition of the `swap matrix':
\begin{align}
\Tr_2 \left[ \mathbb{S} \: \mathbb{1}_{L} \otimes C^{(1)} \right] &= \sum_{i,j=1}^L \Tr_2 \left[ ( E_{i,j} \otimes E_{j,i}) \: (\mathbb{1}_{L}\otimes C^{(1)}) \right]\nonumber \\
&= \sum_{i,j=1}^L \Tr_2 \left[ E_{i,j} \mathbb{1}_{L} \otimes E_{j,i} C^{(1)} \right]\nonumber\\
&= \sum_{i, j=1}^L E_{i,j} \Tr \left[ E_{i,j}^\dagger \: C^{(1)} \right] \nonumber\\
& =\sum_{i, j=1}^L (C^{(1)})_{i,j} E_{i,j} = C^{(1)}.
\label{firstswapID}
\end{align}
We now make use of a result of Ref.~\cite{Elben2019}. Here, it was shown that for general Hilbert spaces $\mathcal{H}$ that there exists a diagonal operator $O^{(1)}$ acting on the doubled (two-copy) Hilbert space $\mathcal{H}^{\otimes 2}$ whose average over the unitary group corresponds to the `swap operator'. Transferred to our situation and in our notation, this reads as follows: Using second order Weingarten calculus of random unitary matrices, we can show (Ref.~\cite{Elben2019}) that there exists a diagonal matrix $O^{(1)} \in \mathcal{M}_L^{\otimes 2}$ with
\begin{align}
 O^{(1)} = \sum_{s_1, s_2=1}^L L(-L)^{\delta_{s_1,s_2}-1} E_{s_1,s_1} \otimes E_{s_2,s_2}.
\label{O1} 
\end{align}
whose average ('twirl') over the unitary group $\mathcal{U}(L)$ is the `swap matrix'
\begin{align}
\Phi^{(2)}(O^{(1)}) &\equiv \int_{\textrm{Haar}}\!\! \!\! \textrm{d}U (U^\dagger)^{\otimes 2} \: O^{(1)} \: (U)^{\otimes 2} \\
&\equiv \overline{(U^\dagger)^{\otimes 2} \: O^{(1)} \: (U)^{\otimes 2}} \nonumber = \mathbb{S}.
\label{eq:snd_order_twirling}
\end{align}
Here, $\int_{\textrm{Haar}}\!\! \textrm{d}U $ denotes the integration over the unitary group $\mathcal{U}(L)$ with respect to the Haar measure, which is abbreviated in the second line by the overline as a short-hand notation. We note that this identity involves only polynomials of unitary transformations $U$ of degree two. Thus, by definition, the Haar integral can be replaced by an average over a {unitary} $2$-design \cite{Collins2010}. 

We insert Eq.~\eqref{eq:snd_order_twirling} and Eq.~\eqref{O1} into Eq.~\eqref{eq:swap_tom} and find
\begin{align}
 &C^{(1)} = \text{Tr}_2\left[ \Phi^{(2)}(O^{(1)}) \mathbb{1} \otimes C^{(1)} \right] \nonumber\\
 &= L\sum_{\substack{s_1, s_2\\=1}}^L (-L)^{\delta_{s_1,s_2}-1} \text{Tr}_2 \overline{\left[ U E_{s_1,s_1}U^\dagger \otimes U E_{s_2,s_2} U^\dagger C^{(1)}\right]}\nonumber\\
 &=L\sum_{\substack{s_1, s_2\\=1}}^L (-L)^{\delta_{s_1,s_2}-1} \overline{ U E_{s_1,s_1} U^\dagger \: \text{Tr}\left[ E_{s_2,s_2}U^\dagger C^{(1)} U\right] } \nonumber \\
 &=L\sum_{\substack{s_1, s_2\\=1}}^L (-L)^{\delta_{s_1,s_2}-1} \overline{ U^T E_{s_1,s_1} U^* \: \text{Tr}\left[ E_{s_2,s_2}U^* C^{(1)} U^T\right] } \nonumber
\end{align}
where we used in the last line can substitute $U$ with $U^T$ without changing the value of the Haar integral. Using now that by definition
\begin{align}
\text{Tr}\left[ E_{s_2,s_2}U^* C^{(1)} U^T\right] 
&= (U^* C^{(1)} U^T)_{s_2,s_2} = N^{(U)}_{s_2}
\end{align}
and specifying to a matrix element $C^{(1)}_{ij}$ yields
\begin{align}
C^{(1)}_{ij} &= L \sum_{s_1,s_2=1}^L 
(-L)^{\delta_{s_1,s_2}-1} \: 
\overline{
  N^{(U)}_{s_2} \:
U_{s_1,i} U^*_{s_1,j}},
\label{1body_sup}
\end{align}
as required.

\subsection{Estimating the two-body correlation matrix}

We now prove Eq.~\eqref{2body} of the main text. In analogy with the one-body problem, we define a correlation tensor $C^{(2)}\in \mathcal{M}^{\otimes 2}_L$ with $L^4$ complex entries
\begin{align}
\label{eq:C2}
(C^{(2)})_{(i,j),(k,l)=1,\dots, L} \ \equiv 
\Tr[\rho  c^\dagger_ic_jc^\dagger_kc_l].
\end{align}
Next, we define a 4-copy swap matrix $\mathbb{S}_{(13)(24)} \in \mathcal{M}^{\otimes 4}_L$ as
\begin{align}
\mathbb{S}_{(13)(24)} \equiv \sum_{i_1, \dots ,j_4} 
E_{i_1,j_3} \otimes E_{i_2,j_4} \otimes E_{i_3,j_1} \otimes E_{i_4,j_2} .
\end{align}
The identity
\begin{align}
C^{(2)} &= \text{Tr}_{3,4} 
\left[ \mathbb{S}_{(13)(24)} \: \mathbb{1}_L \otimes \mathbb{1}_L \otimes C^{(2)} \right].
\label{eq:sndswapid}
\end{align}
can be proven in way similar to Eq.~\eqref{eq:swap_tom}. As in the previous section, one might now ask whether there exists a \emph{diagonal} matrix $O^{(2)} \in \mathcal{M}^{\otimes 4}_L$ which twirls to the 4-copy swap matrix $\mathbb{S}_{(13)(24)}$.  By embedding $C^{(2)}\in \mathcal{M}^{\otimes 2}_{L}$ into a larger tensor $\widetilde{C^{(2)}} \in \mathcal{M}^{\otimes 2}_{L_B}$  defined on a larger lattice $L_B$ by adding zeros
\begin{align}
\widetilde{C^{(2)}}_{i,j,k,l} \equiv \begin{cases}
 C_{i,j,k,l}^{(2)} \quad \textrm{for} \quad 1\leq i,j,k,l \leq L \\
 0 \quad \text{otherwise},
 \end{cases} 
\end{align}
 we will show below that such matrix can indeed be found {approximately}. Concretely, we will show that there exists a diagonal matrix $O^{(2)} \in \mathcal{M}^{\otimes 4}_{L_B}$ such that
\begin{align}
&\left( \text{Tr}_{3,4} 
\left[\Phi^{(4)}(O^{(2)})  \mathbb{1}_{L_B} \otimes  \mathbb{1}_{L_B} \otimes \widetilde{C^{(2)}} \right] \right)_{i,j,k,l}  \nonumber \\
&=\left(C^{(2)} \right)_{i,j,k,l} +\mathcal{O}\left(\frac{L^2}{L_B}\right).
\label{eq:proof1}
\end{align}
Here, $\Phi^{(4)}(O^{(2)})$ denotes the unitary twirling channel of fourth order 
\begin{align}
     \Phi^{(4)}\left(O^{(2)}\right) & \equiv  \int_{\textrm{Haar}}\!\! \!\! \textrm{d}U  (U^\dagger)^{\otimes 4} O  U^{\otimes 4}
\end{align}
and the diagonal matrix $O^{(2)}\in \mathcal{M}^{\otimes 4}_{L_B}$ is defined as \begin{align*}
    O^{(2)}= \! \! \sum_{\substack{s_1,s_2,\\s_3,s_4}=1}^{L_B} \! \! o^{(2)}_{s_1,s_2,s_3,s_4}  E_{s_1,s_1} \otimes E_{s_2,s_2}  \otimes E_{s_3,s_3} \otimes E_{s_4,s_4}.
\end{align*}
with coefficients  $o^{(2)}_{s_1,s_2,s_3,s_4}$ as stated in the Eq.~\eqref{coeff2body} of the main text. Inserting these definitions, we find that the left-hand-side of Eq.~\eqref{eq:proof1} indeed equals to the right-hand-side of Eq.~\eqref{2body} of the main text:
\begin{align}
&\left( \text{Tr}_{3,4} 
\left[\Phi^{(4)}(O^{(2)})  \mathbb{1}_{L_B} \otimes  \mathbb{1}_{L_B} \otimes \widetilde{C^{(2)}} \right] \right)_{i,j,k,l}  \nonumber \\
&=\sum_{\substack{s_1,s_2,\\s_3,s_4}=1}^{L_B} o^{(2)}_{s_1,s_2,s_3,s_4}  \overline{
N^{(U)}_{s_3,s_4} \: 
U_{s_1,i} U^*_{s_1,j} U_{s_2,k} U^*_{s_2,l}} .
\end{align}
Thus, it remains to show that Eq.~\eqref{eq:proof1} holds true.

\subsubsection{Proof of  Eq.~\eqref{eq:proof1} }

The key ingredient in our proof is the  unitary twirling channel  $\Phi^{(4)}$.  Exploiting Schur-Weyl duality,  it can be derived  \cite{Roberts2017} that, $\Phi^{(4)}$  evaluates to 
\begin{align}
      \Phi^{(4)}\left(O^{(2)}\right) &=  \int_{\textrm{Haar}}\!\! \!\! \textrm{d}U  (U^\dagger)^{\otimes 4}O^{(2)} U^{\otimes 4} \nonumber \\
    & = \sum_{\pi,\sigma \in S_4} C_{\pi,\sigma} W_{\pi} \Tr[W_\sigma O^{(2)} ].
    \label{eq:twirling4}
\end{align}
Here $S_4$ is the permutation group of four items, composed by $4!=24$ elements, $C_{\pi,\sigma}$ is the real-valued and symmetric Weingarten matrix \cite{Collins2003} and $W_\pi$ is a permutation matrix related to the element $\pi \in S_4$, defined as
\begin{align}
    W_\pi = \sum_{\substack{s_1,s_2,\\s_3,s_4}=1}^{L_B} E_{s_{\pi(1)},s_1} \otimes E_{s_{\pi(2)},s_2} \otimes E_{s_{\pi(3)},s_3}\otimes E_{s_{\pi(4)},s_4}.
    \label{eq:Wpi}
\end{align}
Note that in equation~\eqref{eq:twirling4}, the `Gaussian terms' correspond to the terms of equal permutations $\sigma=\tau$ that, as we show below,  become dominant for our choice of $O^{(2)}$ in the thermodynamic limit $L_B\to\infty$~(see also \cite{Enk2012}).
We will assume for the moment that it exists an operator $O^{(2)}$ that fulfills
\begin{align}
\Tr[W_\pi O^{(2)}] &= C^{-1}_{\pi \pi} \quad \text{for} \quad W_\pi = \mathbb{S}_{(13)(24)} \label{cond_1}\\
\Tr[W_\pi O^{(2)}] &= 0 \quad \text{otherwise}.
\label{cond_2}
\end{align} 
Using this result, that we will prove in the next section, the action of the twirling channel greatly simplifies. Denoting the permutation $\gamma\equiv (13)(24)\in S_4$, we find
\begin{align}
\Phi^{(4)}(O^{(2)}) &= \sum_{\pi,\sigma \in S_4} C_{\pi,\sigma} W_{\pi} \Tr[W_\sigma O^{(2)}] \nonumber\\
&= \sum_{\pi\in S_4} \frac{C_{\pi,\gamma}}{C_{\gamma,\gamma}} W_{\pi} \nonumber \\
&=\mathbb{S}_{(13)(24)} + \sum_{\pi\in S_4, \pi \neq \gamma} \frac{C_{\gamma,\pi}}{C_{\gamma,\gamma}}W_{\pi} 
\label{OL},
\end{align}
where we have separated the Gaussian (diagonal) terms from the non-Gaussian (off-diagonal) terms. 
We plug this into Eq.~\eqref{eq:proof1} and find
\begin{align}
&\left( \text{Tr}_{3,4} 
\left[\Phi^{(4)}(O^{(2)})  \mathbb{1}_{L_B} \otimes  \mathbb{1}_{L_B} \otimes \widetilde{C^{(2)}} \right] \right)_{i,j,k,l}  \nonumber \\
&=\left( \text{Tr}_{3,4} 
\left[ \mathbb{S}_{(13)(24)}   \mathbb{1}_{L_B} \otimes  \mathbb{1}_{L_B} \otimes \widetilde{C^{(2)}} \right] \right)_{i,j,k,l}  \nonumber \\
&\quad + \sum_{\pi\in S_4, \pi \neq \gamma}    \frac{C_{\gamma,\pi}}{C_{\gamma,\gamma}}  \left(\text{Tr}_{3,4}  \left[ W_{\pi}  \mathbb{1}_{L_B} \otimes  \mathbb{1}_{L_B} \otimes \widetilde{C^{(2)}} \right] \right)_{i,j,k,l} \nonumber \\&
=\left(C^{(2)} \right)_{i,j,k,l}    \label{eq:prooftwo} \\
&\quad + \sum_{\pi\in S_4, \pi \neq \gamma}    \frac{C_{\gamma,\pi}}{C_{\gamma,\gamma}}  \left( \text{Tr}_{3,4} \left[W_{\pi}    \mathbb{1}_{L_B} \otimes  \mathbb{1}_{L_B} \otimes \widetilde{C^{(2)}} \right] \right)_{i,j,k,l} . \nonumber
\end{align}
While the first Gaussian term in Eq.~\eqref{eq:prooftwo} has the desired form, we need to bound the non-Gaussian off-diagonal corrections. Using the triangle inequality, we find
\begin{align}
    &\sum_{\pi\in S_4, \pi \neq \gamma}    \frac{C_{\gamma,\pi}}{C_{\gamma,\gamma}}  \left(\text{Tr}_{3,4}  \left[W_{\pi}    \mathbb{1}_{L_B} \otimes  \mathbb{1}_{L_B} \otimes \widetilde{C^{(2)}} \right] \right)_{i,j,k,l}  \\ &\leq  \sum_{\pi\in S_4, \pi \neq \gamma}  \left| \frac{C_{\gamma,\pi}}{C_{\gamma,\gamma}} \right| \left| \left( \text{Tr}_{3,4} \left[W_{\pi}    \mathbb{1}_{L_B} \otimes  \mathbb{1}_{L_B} \otimes \widetilde{C^{(2)}} \right] \right)_{i,j,k,l}  \right| \nonumber.
\end{align}
Using the asymptotic expansion of the Weingarten function for large $L_B$ \cite{Collins2010}, we find that for $\gamma \neq \pi$ it decreases as
\begin{align}
   \left| \frac{C_{\gamma,\pi}}{C_{\gamma,\gamma}} \right|  = \mathcal{O}\left(L_B^{-1}\right) \quad \text{for}\quad \pi \neq \gamma.
\end{align}
Finally, we show that the second factors can be bounded by a polynomial in $L$. We insert the definition of $W_\pi $ [Eq.~\eqref{eq:Wpi}] to show that
\begin{align}
&\left|\left( \text{Tr}_{3,4}  \left[W_{\pi}    \mathbb{1}_{L_B} \otimes  \mathbb{1}_{L_B} \otimes \widetilde{C^{(2)}} \right] \right)_{i,j,k,l} \right| \\
& \leq \sum_{\substack{s_1,s_2,\\s_3,s_4}=1}^{L_B}  \delta_{i,s_{\pi(1)}}\delta_{j,s_{1}} \delta_{k,s_{\pi(2)}}\delta_{l,s_{2}} |\widetilde{C^{(2)}} _{s_3,s_{\pi(3)},s_4,s_{\pi(4)}}| \nonumber.
\end{align}
For fixed indices $i,j,k,l$, this sum contains at most $L^2$ terms of the form $|\widetilde{C^{(2)}} _{s_3,s_{\pi(3)},s_4,s_{\pi(4)}}|=\mathcal{O}(1)$ (as $\widetilde{C^{(2)}}$ for a fermionic system contains only $L^2$ non-vanishing terms of $\mathcal{O}(1)$). For a bosonic system, where all particles can coexist on the same site, $|\widetilde{C^{(2)}}_{s_3,s_{\pi(3)},s_4,s_{\pi(4)}}|=\mathcal{O}(N^2)$.
Since here we are dealing with fermionic states, we can bound the corrections in Eq.~\eqref{eq:prooftwo} as 
\begin{align}
&\sum_{\pi\in S_4, \pi \neq \gamma}    \frac{C_{\gamma,\pi}}{C_{\gamma,\gamma}}  \left( \text{Tr}_{3,4} \left[W_{\pi}    \mathbb{1}_{L_B} \otimes  \mathbb{1}_{L_B} \otimes \widetilde{C^{(2)}} \right] \right)_{i,j,k,l} \nonumber \\ & = \mathcal{O}\left(\frac{ L^2}{L_B}\right).
\end{align}
This shows Eq.~\eqref{eq:proof1}. In App.~D, we also show numerically how the corrections, giving rise to systematic error (bias) of the estimator, decay with $L_B$.

\subsubsection{Deriving the coefficients of $O^{(2)}$}

The aim of this subsection is to find the coefficients $o^{(2)}_{s_1,s_2,s_3,s_4}$ of  $O^{(2)}$
such that it fulfills the desired relations Eqs.~\eqref{cond_1} and \eqref{cond_2}. The solution of this problem is not unique. We thus proceed in two steps. We first make an explicit ansatz and derive a solution. Secondly, we symmetrize over many possible choices with the same structure. We show numerically that this symmetrization decreases statistical errors. 

First, we start by considering the following ansatz:
\begin{align}
\label{ABCD}
o_{s_1,s_2,s_3,s_4} = \frac{1}{\mathcal{L}} A_{s_1} B_{s_2} A_{s_3} B_{s_4}
\end{align}
where $\mathcal{L}=C_{\gamma,\gamma}^{-1}=\prod_{i=0}^3 (L_B+i)$ is a constant and $A$, $B$, $C$ and $D$ are vectors made of $L_B$ elements defined as: 
\begin{align}
A_s &= 
\frac{1}{\mathcal{N}_A}
\begin{cases} 
 1 & s\ \text{odd}  \\
-1 & s\ \text{even} \\
 0 & s>L_B/2 \\
\end{cases}\\
B_s &=
\frac{1}{\mathcal{N}_B}
\begin{cases} 
 0 & s<L_B/2 \\
 1 & s\ \text{odd} \\
-1 & s\ \text{even} 
\end{cases}.
\end{align}
Here the normalization constant $\mathcal{N}_A$ (and similarly for $\mathcal{N}_B$) is defined as $\mathcal{N}_A=\sqrt{\sum_{i=1}^{L_B} |A_i|^2}$.
Explicitly, $A$ and $B$, for $L_B=8$ take then the form:
\begin{align}
A &= 
\frac{1}{\mathcal{N}_A}
\begin{pmatrix} 1 & -1 & 1 & -1 & \dots & 0 & 0 & 0 & 0 \end{pmatrix} \\
B &= 
\frac{1}{\mathcal{N}_B} 
\begin{pmatrix} 0 & 0 & 0 & 0 & \dots & 1 & -1 & 1 & -1 \end{pmatrix}.
 \label{explicit}
\end{align}
Since $A$ and $B$ are composed by $L_B/2$ non-zero elements, then $\mathcal{N}_A = \mathcal{N}_B = \sqrt{L_B/2}$. 
We can now verify that this ansatz fulfills \eqref{cond_1} and \eqref{cond_2}. First, we note that:
\begin{align}
\sum_{s=1}^{L_B} A_s = \sum_{s=1}^{L_B} B_s = \sum_{s=1}^{L_B} A_s B_s &= 0 \\
\sum_{s=1}^{L_B} A_s^2 B_s = \sum_{s=1}^{L_B} A_s B_s^2 =
\sum_{s=1}^{L_B} A_s^2 B_s^2 &= 0  \\
\sum_{s=1}^{L_B} A_s^2 = \sum_{s=1}^{L_B} B_s^2  &= 1 .
\label{contr}
\end{align}
By direct calculation using the previous relations \eqref{contr}, one finds that it holds
\begin{align}
\Tr[\mathbb{S}_{(13)(24)} O^{(2)}] &=
\frac{1}{\mathcal{L}}
(\sum_{s=1}^{L_B} A_s^2 )
(\sum_{p=1}^{L_B} B_p^2 ) = \frac{1}{\mathcal{L}} \\
\Tr[W_\pi O^{(2)}] = 0 \ \ \ \ & \forall \ \ W_\pi \neq \mathbb{S}_{(13)(24)} .
\end{align}

Secondly, we symmetrize the above ansatz by noticing that any permutation of the elements of A and B in Eq.~\eqref{ABCD} also fulfils the conditions \eqref{cond_1} and \eqref{cond_2}.
We therefore define a $\tilde{O}^{(2)}$ from averaging over all the permutations of $L_B$ elements as:
\begin{align}
\tilde{o}_{s_1 \dots s_4} = 
\frac{1}{\mathcal{L}}
\sum_{\sigma \in S_{_B}} 
\frac{1}{L_B!}
A_{\sigma(s_1)}
B_{\sigma(s_2)}
A_{\sigma(s_3)}
B_{\sigma(s_4)} .
\end{align}

\begin{figure}[h!!!!!!t]
 \centering
 \includegraphics[width=0.7\columnwidth]{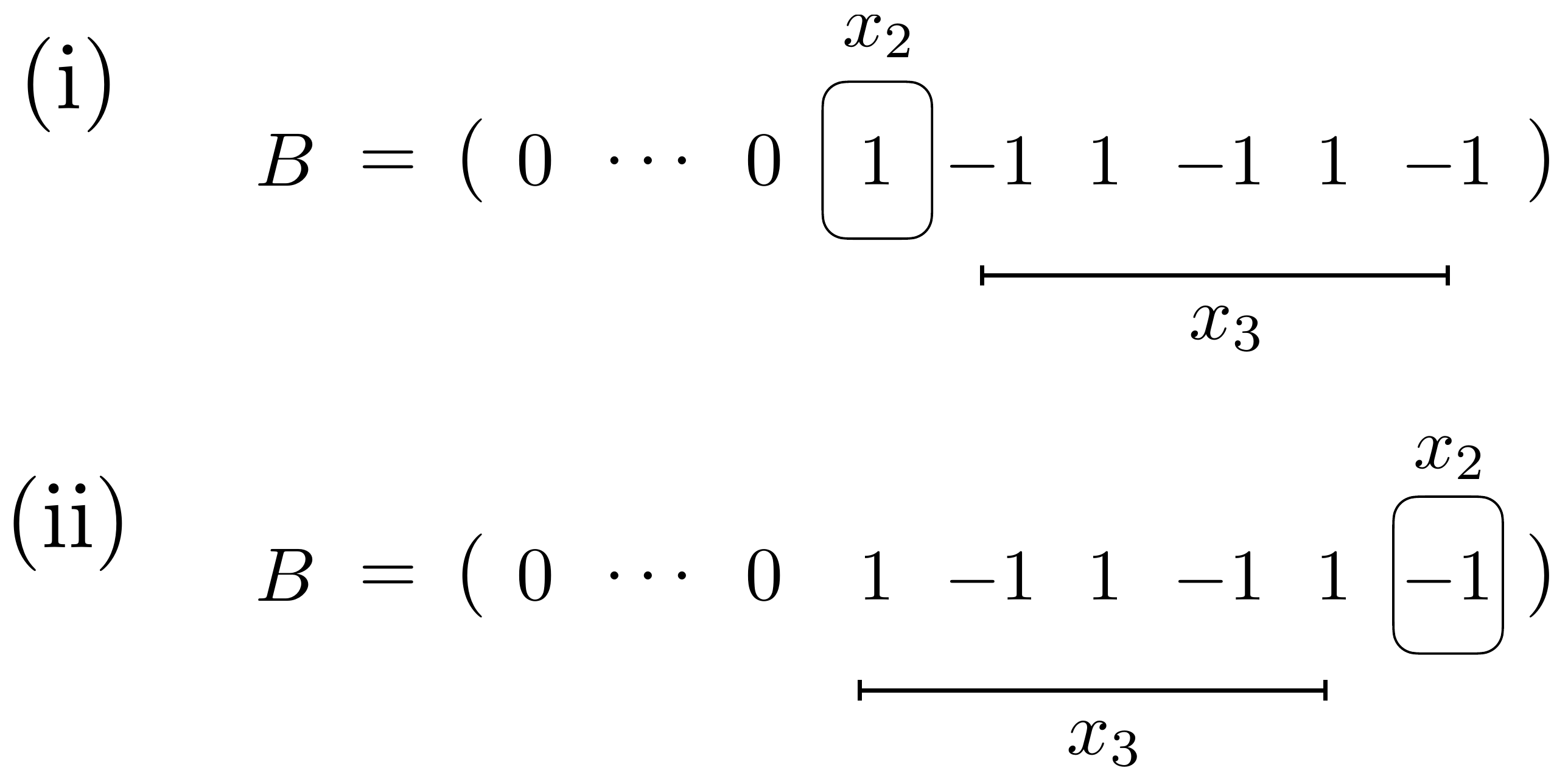}
 \caption{Schematic representation of the two possibles cases of $B_{x_2} B_{x_3}$.}
 \label{fig:C_1}
\end{figure}

Given the structure of the ansatz \eqref{ABCD}, and the explicit form of $A$ and $B$ given in \eqref{explicit} we can make the following statement. For any permutation, a necessary but not sufficient condition to have a non zero element in the sum is:
\begin{align}
& \left( \sigma(s_1) \neq \sigma(s_2) \right) \wedge 
  \left( \sigma(s_1) \neq \sigma(s_4) \right) \ \ \ \text{and} \\
& \left( \sigma(s_3) \neq \sigma(s_2) \right) \wedge 
  \left( \sigma(s_3) \neq \sigma(s_4) \right) .
\end{align}
Since each permutation is a one-to-one function, this condition corresponds to:
\begin{align}
& \left( s_1 \neq s_2 \right) \wedge 
  \left( s_1 \neq s_4 \right) \ \ \ \text{and} \\
& \left( s_3 \neq s_2 \right) \wedge 
  \left( s_3 \neq s_4 \right) .
\end{align}
All the elements fulfilling this condition can be separated in three categories: 1) all four indices different, 2) one couple of equal indices, 3) two different couples of equal indices. 
$\tilde{o}_{s_1 \dots s_4}$ can be then expressed in a condition form:
\begin{align}
\tilde{o}_{s_1 \dots s_4} &= 
\begin{cases} 
C_0 &
s_1 \! \neq \! s_2 \! \neq \! s_3 \! \neq \! s_4 \\[3pt]
C_1 &
\left( s_1 \! = \! s_3 \right) \:\: \dot{\lor} \:\:
\left( s_2 \! = \! s_4 \right) \\[3pt]
C_2 & s_1 \! = \!s_3 \! \neq \! s_2 \! = \!s_4 \\[3pt]
0 &\text{otherwise} .
\end{cases}
\label{cond}
\end{align}
These three coefficients can be obtained by combinatory calculations, carrying out the sum over all the permutations.
Let us first consider the case of $C_2$. Since the same permutation is applied to the elements of both vectors $A$ and $B$, the condition \eqref{cond}, for $C_2$, can be recast into: 
$s_1 \! = \! s_3 \! \neq \! s_2 \! = \! s_4$. We therefore want to calculate
\begin{align}
C_2 = \frac{1}{\mathcal{L}} \frac{1}{L_B!}
\sum_{\sigma \in S_{L_B}} 
A_{\sigma(s_1)}^2
B_{\sigma(s_2)}^2 \ \ \ s_1 \neq s_2 .
\label{C2}
\end{align}
Let us define $P_{nz}$ as the set of the $(L_B/2)^2$ pairs of indices $\{x_1,x_2\neq x_1\}$ for which it holds $A_{x_1}^2 B_{x_2}^2 = 1$. Only the permutations for which $\{\sigma(s_1),\sigma(s_2)\} \in P_{nz}$ add a non zero contribution to the sum \eqref{C2}. For each element in $\{ x_1,x_2 \} \in P_{nz}$, there are $(L_B\!-\!2)!$ permutations that leave unchanged the elements $A_{x_1}$ and $B_{x_2}$. The number of non-zero terms in the sum is, therefore $(L_B/2)^2 (L_B\!-\!2)!$. This leads us to the result:
\begin{align}
C_2 &= \frac{1}{\mathcal{L}} \frac{1}{L_B!}
\frac{1}{\mathcal{N}_A^2}
\frac{1}{\mathcal{N}_B^2}
 \left(
\frac{L_B}{2}
\right)^2 
(L_B-2)! \\
& = \frac{(L_B-2)!}{\mathcal{L}\: L_B!} .
\end{align}
The calculation of the coefficient $C_1$ is more involved. Here only one couple of indices are allowed to be equal, either $s_1=s_3$ or $s_2=s_4$. Without loss of generality let's assume $s_1=s_3$.
\begin{align}
C_1 = \frac{1}{\mathcal{L}} \frac{1}{L_B!}
\sum_{\sigma \in S_{L_B}} 
A_{\sigma(s_1)}^2
B_{\sigma(s_2)} B_{\sigma(s_4)} \ \ \ s_1 \neq s_2 \neq s_4 .
 \label{C_1}
\end{align}
In analogy with the previous case we define $T_{nz}$ as the set of the $(L_B/2)^2(L_B/2\:-\:1)$ triplets of different indices $\{x_1,x_2,x_3\}$ for which it holds $A_{x_1}^2 B_{x_2} B_{x_3} \neq 0$. It must be noticed that for any element in $T_{nz}$, $A_{x_1}^2=1$.
For any triplet in $T_{nz}$, given a choice of $x_1$, there are two cases illustrated in Fig.~\ref{fig:C_1}.

\begin{figure}[h!!!!!!]
 \centering
 \includegraphics[width=0.75\columnwidth]{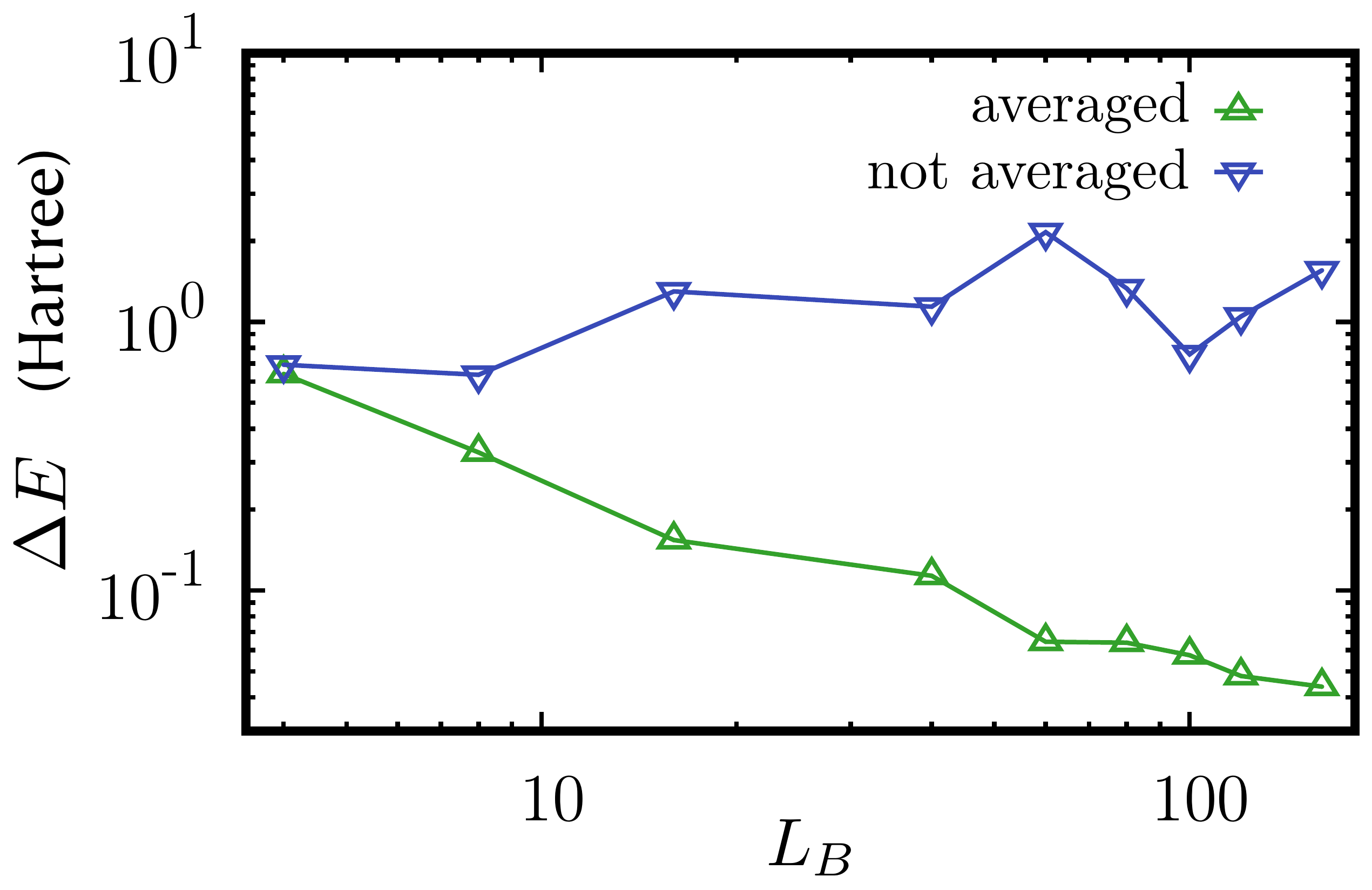}
 \caption{Scaling of statistical error $\Delta E$ for the Li-H molecule studied in the main text, as a function of $L_B$. Numerical results for both $O^{(2)}$ ansatz, not averaged in Eq.~\eqref{ABCD} and averaged in Eq.~\eqref{cond}. Here we assume $N_m=\infty$. $N_u=10^3$ unitary transformations have been considered and the bond distance is set to $1.5\:\si{\angstrom}$.}
 \label{fig:aver}
\end{figure}

If $B_{x_2}\!=\!1$, then then choice of $x_3$ in $L_B/4$ cases leads $A_{x_1}^2 B_{x_2} B_{x_3}=-1$, while in the others $L_B/4-1$ it leads to $A_{x_1}^2 B_{x_2} B_{x_3}=1$. Repeating now the reasoning used for $C_2$, in the sum \eqref{C_1}, we will have a total $(L_B/2)^2(L_B/4\:-\:1)$ positive term and $(L_B/2)^2(L_B/4)$ negatives terms. Summing up these two contributions we will get:
\begin{align}
C_1 &= - \frac{1}{\mathcal{L}} \frac{1}{L_B!}
\frac{1}{\mathcal{N}_A^2}
\frac{1}{\mathcal{N}_B^2}
\left(\frac{L_B}{2}\right)^2
 (L_B-3)! \\
 &= - \frac{1}{\mathcal{L}} \frac{(L_B-3)!}{L_B!} .
\end{align}
A similar procedure can be applied to obtain $C_0$.
\begin{align}
C_0 &= \frac{1}{\mathcal{L}}
\frac{(L_B-4)!}{L_B!} . 
\end{align}
We now study numerically the impact of the two ansatzes, not averaged \eqref{ABCD} and averaged \eqref{cond}, on the reconstruction of $C^{(2)}$ in equation \eqref{eq:C2}. Here we take in consideration the case of the Li-H molecule we discuss in the main text. As show in Fig.~\ref{fig:aver} while increasing $L_B$ the averaged ansatz helps to strongly suppress statistical errors.

\section{Appendix D: ANALYSIS OF SYSTEMATIC ERRORS FOR LI-H MOLECULE}
\label{app:D}
\begin{figure}[h!!!!!!]
 \centering
 \includegraphics[width=\columnwidth]{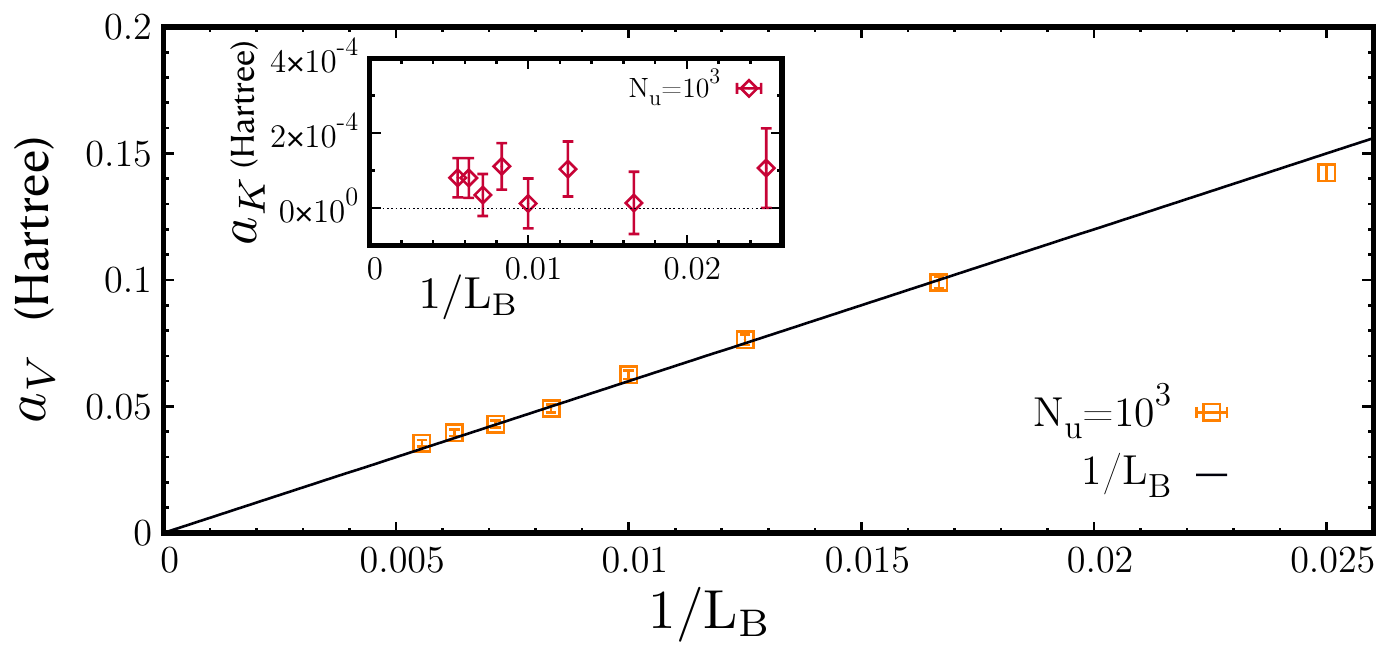}
 \caption{Reconstruction error of both kinetic (inset) and potential energy as a function $L_B^{-1}$, for $N_u=10^3$ unitary transformations. Bond distance is set to $1.5\:\si{\angstrom}$. We assume here $N_m=\infty$.}
 \label{fig:misc_2}
\end{figure}

In this last appendix we study numerically the systematic error we introduce in estimating two-body correlations $C^{(2)}_{ijkl}$ by considering a finite value of $L_B$.
This effect is shown in Fig.~\ref{fig:misc_2} for the Li-H molecule ($N=2$, $L=4$), where we represent the accuracy $a_K=|\tilde K-K|$ in estimating the kinetic part $K$ of the molecular Hamiltonian, as well as the accuracy $a_V=|\tilde V-V|$ for the interacting part related to the two-body correlations. Here, $\tilde K$ ($\tilde V$) denote the estimation provided by the protocol of the kinetic (interaction respectively) part of the molecular energy. As seen in this example, the estimation of the interaction $V$ becomes more accurate as the size of the extended system $L_B$ increases. The kinetic part that relies only on single body correlations is, on the other hand, accurate for any $L_B$. 

\clearpage
\end{document}